\documentclass[reprint,onecolumn]{revtex4-2}
\usepackage[latin9]{inputenc}   
\usepackage{mathtools}
\usepackage{amstext}
\usepackage{amsthm}
\usepackage{graphicx}

\makeatletter  

\newcommand{\lyxdot}{.}

\numberwithin{equation}{section}
\numberwithin{figure}{section}
\theoremstyle{plain}
\newtheorem{thm}{\protect\theoremname}
\theoremstyle{plain}
\newtheorem{lem}[thm]{\protect\lemmaname}
\theoremstyle{plain}
\newtheorem{cor}[thm]{\protect\corollaryname}

\usepackage{dcolumn}
\usepackage{amsfonts} 
\usepackage{xcolor}

\newcommand{\adj}{^\dagger}
\newcommand{\ddt}[1]{\frac{ \textrm{d} #1}{ \textrm{d} t}}

\newcommand{\onov}[1]{\frac{1}{#1}}

\newcommand{\mc}[1]{\mathcal{#1}}
\newcommand{\R}{\mathbb{R}}
\newcommand{\C}{\mathbb{C}}
\newcommand{\Z}{\mathbb{Z}}
\newcommand{\N}{\mathbb{N}}
\newcommand{\wif}{\quad \textrm{if} \quad}

\newcommand{\Li}{\textrm{Li}}

\@ifundefined{showcaptionsetup}{}{%
 \PassOptionsToPackage{caption=false}{subfig}}
\usepackage{subfig}
\makeatother  

\providecommand{\corollaryname}{Corollary}
\providecommand{\lemmaname}{Lemma}
\providecommand{\theoremname}{Theorem}

\begin{document}
\title{Time and space generalized diffusion equation on graph/networks}
\author{Fernando Diaz-Diaz and Ernesto Estrada}
\address{Institute of Cross-Disciplinary Physics and Complex Systems, IFISC
(UIB-CSIC), 07122 Palma de Mallorca, Spain}
\begin{abstract}
Normal and anomalous diffusion are ubiquitous in many complex systems \cite{Diffusive_Bunde_2018}.
Here, we define a time and space generalized diffusion equation (GDE),
which uses fractional-time derivatives and transformed $d$-path Laplacian
operators on graphs/networks. We find analytically the solution of
this equation and prove that it covers the regimes of normal, sub-
and superdiffusion as a function of the two parameters of the model.
We extend the GDE to consider a system with temporal alternancy of
normal and anomalous diffusion which can be observed for instance
in the diffusion of proteins along a DNA chain. We perform computational experiments
on a one-dimensional system emulating a linear DNA chain. It is shown
that a subdiffusive-superdiffusive alternant regime allows the diffusive
particle to explore more slowly small regions of the chain with a
faster global exploration, than a subdiffusive-subdiffusive regime.
Therefore, an alternancy of sliding (subdiffusive) with hopping and
intersegmental transfer (superdiffusive) mechanisms show important
advances for protein-DNA interactions.
\end{abstract}

\maketitle

\section{Introduction}

Diffusion--the net movement of particles in an environment, generally
from a region of higher concentration to a region of lower concentration--is
ubiquitous in natural and man-made systems. In the absence of obstacles
to diffusion and traps, the diffusive particles describe a random
walk motion on the environment, such that their mean squared displacements
(MSD) scale linearly with time, $\langle x^{2}\rangle\propto t$.
This process is known as normal diffusion. However, the existence
of obstacles in the environment may trigger long-jumps of the diffusive
particle  \cite{chen_mathematical_2020, king_non-local_2021, yu_single-molecule_2013}, such that the mean displacement of the particles is bigger
than that of the normally diffusing ones in the same period of time,
i.e., $\langle x^{2}\rangle\propto t^{\gamma>1}$. This type of process
is known as superdiffusive. On the other hand, it is possible that
the environment has regions acting as traps for the particles, where
they are retained for longer times than in a normal diffusive process.
In this case, $\langle x^{2}\rangle\propto t^{\gamma<1},$ and the
process is known as subdiffusive. From the modeling perspective there
are several approaches to describe anomalous (sub- and super-) diffusion
\cite{metzler_random_2000,sokolov2012models}. From a physical perspective
these processes have analogues in terms of anomalous heat conduction
\cite{li2003anomalous}, where normal diffusion implies normal heat
conduction, superdiffusion implies anomalous heat conduction with
a divergent thermal conductivity and subdiffusion implies anomalous
heat conduction with a convergent thermal conductivity. In the last
case the system is a thermal insulator in the thermodynamic limit.

It has been remarked that subdiffusion may arise as the result of
the coexistence of time-periods dominated by normal transport with
periods in which there is no effective transport. The last can emerge
when the diffusive particle is temporarily trapped as a result of
geometrical complexity and interactions with the environment. This
could be clearly the case of the travel of contaminants in groundwater,
which display much longer times than the ones expected from the classic
diffusion. The motion of proteins while sliding on DNA during target
search is believed to be subdiffusive in general \cite{kong2017rad4}.
Simulations results for the case of T7 RNA polymerase promoter search
on T7 DNA has been found to be subdiffusive for short times and asymptotically
approaching normal diffusion \cite{barbi2004model}. More recent intensive
computational simulations also pointed out the important role of subdiffusive
process in the diffusive search of proteins for their specific binding
sites on DNA in the presence of the macromolecular crowding in cells
\cite{liu2017facilitated}. Nowadays it is well established that the
molecular crowding of the internal cellular environment induces the
emergence of anomalous subdiffusion of cytoplasmic macromolecules.
This has been verified by means of fluorescence correlation spectroscopy
and computer simulations \cite{weiss2004anomalous}, fluorescence
correlation spectroscopy \cite{banks2005anomalous}, and by tracking
fluorescently labeled mRNA molecules \cite{golding2006physical}.
The complexities of the process where studied with globular proteins
dispersed in aqueous solution of poly(ethylene oxide) (PEO) to mimic
a crowded environment. Using state-of-the-art neutron spin echo (NSE)
and small-angle neutron scattering (SANS) techniques it was observed
a fast dynamic corresponding to diffusion inside a trap built by the
polymer mesh with slower process corresponding to the long time diffusion
on macroscopic length scales \cite{gupta2016protein}. It has also been found that water molecules jump randomly between trapping sites on protein surfaces, giving rise to subdiffusion. At longer times the subdiffusive exponent gradually increases towards normal diffusion due to a many-body volume-exclusion effect \cite{tan2018gradual}.

The intermittency of fast and slow processes as the one described
by the globular proteins in a crowded environment \cite{gupta2016protein}
can also be found in other scenarios. For instance, it has been reported
that a proliferating, diffusing tumor within different surrounding
tissue conditions migrates not only by using normal diffusion, but
also using combinations of subdiffusion, superdiffusion, and even
ballistic diffusion, with increasing mobility of the tumor cell when
haptotaxis and chemotaxis toward the host tissue surrounding the proliferative
tumor are involved \cite{jiang2014anomalous}. The cytoskeleton (CSK), a
crowded network of structural proteins that stabilizes cell shape
and drives cell motions, displays spontaneous subdiffusive bead motions
at short times followed by superdiffusive motion at longer times.
The intermittency of the motions depended on both the approach to
kinetic arrest and energy release due to ATP hydrolysis \cite{bursac2005cytoskeletal}. 

If we consider the protein\textquoteright s diffusive transport on
DNA in a wider perspective, i.e., not only considering the sliding
process, then we observe a whole range of normal and anomalous behavior.
Using recent advances in single molecule detection it has been observed
that proteins diffusing along DNA follow different mechanisms, such
as (i) random collision, (ii) sliding, (iii) hopping, (iv) intersegmental
transfer and (v) active translocation \cite{shimamoto1999one,gorman2008visualizing}.
While the sliding can give rise to subdiffusive and normal diffusion
behavior, the hopping is known in other systems to produce superdiffusive
behavior \cite{super_1,super_2,super_3,super_4}. Additionally, intersegmental
transfers can transport a protein from one site in the DNA to another
very distant from the original one \cite{inter_1,inter_2,inter_3},
which can give rise to superdiffusive behavior. 

Many of the complex systems in which these normal and anomalous diffusion
processes take place form interaction networks \cite{boccaletti_complex_2006, estrada_structure_2011}. For instance, DNA
can be represented as a linear chain on which a protein is diffusing
between its nodes and edges. Therefore, here we consider the time
and space generalization of the diffusion equation on graphs/networks.
We consider time-fractional derivatives, which account for nonlocality
by time or dynamic memory \cite{tarasov2018no,du2013measuring}, and
long-range jumps in the graph/network through the $d$-path Laplacian
operators \cite{estrada_path_2012,estrada_path_2017, estrada_path_2018}. We first define
the generalized diffusion equation (GDE), study the main properties
of its solution and find analytically the conditions for the existence
of normal, sub- and superdiffusion. We also consider a time-varying
GDE such that the three diffusive regimes appear intermittently with
time. Finally, we apply this approach to the study of the diffusion
of a particle through a linear chain representing a protein diffusing
through DNA.

\section{Preliminaries}

Here we will use interchangeably the terms graph and network for $G=(V,E),$
which in general will be connected and undirected. If the number of
nodes $n=\#V$ is infinite we then assume that $G$ is locally finite
(all vertices have finite degree). Furthermore, let $l^{2}(V)$ be
the Hilbert space of square-summable functions on $V$. 

We begin by defining the $d$-path Laplacian operator on $G$. Let
$dist(v,w)$ be the length of the shortest path between $v$ and $w$,
and let $k_{d}(v)$ be the $d$-path degree of the vertex $v$, defined
by:
\begin{align}
k_{d}(v)=\#\{w\in V:dist(v,w)=d\}.
\end{align}
Let $f$ be a function acting over the set of vertices of $G.$ Then,
the $d$-path Laplacian operator $L_{d}$ is defined by: 
\begin{align}
(L_{d}f)(v)=\sum_{w\in V:dist(v,w)=d}(f(v)-f(w)).
\end{align}
 Let $e_{v}(w)$ be the orthonormal basis 
\begin{align}
e_{v}(w)=\begin{cases}
1\quad\textrm{if}\quad w=v,\\
0\quad\textrm{otherwise.}
\end{cases}\label{orthonormal_basis}
\end{align}

Then, we have 
\begin{align}
(L_{d}e_{v})(w)=\begin{cases}
k_{d}(v)\quad\ \textrm{if}\quad w=v,\\
-1\ \ \ \ \quad\textrm{if}\quad dist(v,w)=d,\\
0\quad\ \ \ \ \ \ \textrm{otherwise.}
\end{cases}
\end{align}
The Mellin-transformed $d$-path Laplacian is the following weighed
sum of $d$-path Laplacians: 
\begin{align}
\tilde{L}(s)=\sum_{d=1}^{\Delta}L_{d}d^{-s}, \label{MTDPL}
\end{align}
where $\Delta$ is the diameter of the graph. This operator preserves
several key properties of the graph Laplacian \cite{estrada_path_2012, estrada_path_2017}:
\begin{lem}
The Mellin-transformed d-path Laplacian is positive semidefinite.
Furthermore, let $k_{d,max}:=max\{k_{d}(v):v\in V\}$. If $k_{d,max}\leq Ck^{\alpha}$,
then $\tilde{L}(s)$ is bounded for all $s\in\C$ with $\mathbb{R}(s)>\alpha+1$. 
\end{lem}

When $s\rightarrow\infty$, all non-unity entries of $\tilde{L}(s)$
vanish and we have the standard graph Laplacian operator $L$. The physical difference of the d-path Laplacian operator with the fractional powers of the standard Laplacian have been analyzed in \cite{estrada_path_2021}. The standard diffusion equation on a graph is formulated on the basis
of this operator as follows:
\begin{align}
\frac{\partial f(t)}{\partial t}=-\mathcal{D}Lf(t), & f\left(0\right)=f_{0},
\end{align}
where $\mathcal{D}$ is the diffusion coefficient, hereafter taken
always to be unity. This equation has the well-known solution: 
\begin{align}
f(t)=e^{-Lt}f_{0} & .
\end{align}
The function $f(t)$ can be interpreted as a probability density function (pdf) of the position of a hypothetical diffusing particle. Indeed, the diffusion equation ensures that several key properties of a pdf are satisfied: (i) If $(f_0)_n\geq0 \ \forall n$, the exponential operator $e^{-Lt}$ ensures that the time evolved pdf remains positive: $(f(t))_n\geq0 \ \forall n, t$. (ii) If the initial condition is normalized, ($\sum_n (f_0)_n=1$) then $f(t)$ remains normalized for any $t$.  This is because the diffusion equation preserves the 1-norm $\parallel f(t) \parallel_1:=\sum_n f_n(t)$:
\begin{align}
    \ddt{\parallel f(t) \parallel_1} = \ddt{ } \left( \sum_i f_i  \right ) =-\sum_i (Lf)_i=-\sum_{j} L_{jj} f_j -\sum_{i,j\neq i} L_{i j} f_j = \sum_{i,j\neq i} L_{ij} f_j -\sum_{i,j\neq i} L_{ij} f_j =0,
\end{align}
where we have used $L_{jj}=-\sum_{i\neq j} L_{i j}$. The above equation implies that $\parallel f_i\parallel _1(t)$ is constant, as previously stated.  \\
When the diffusion equation is defined as before, the mean square
displacement (MSD) $\langle x^{2}\rangle$ of the diffusive particle
scales linearly with time: $\langle x^{2}\rangle\propto t$, independently
of the initial conditions and the diffusion coefficient. This is
known as normal diffusion in the literature. However, many physical
systems display the so-called \textit{anomalous diffusion} \cite{metzler_random_2000, sokolov2012models}, where
the MSD scales as a power law with time: $\langle x^{2}\rangle\propto t^{\gamma}$.
If $\gamma<1$, the system is in a subdiffusive regime, whereas if
$\gamma>1$, the system is in a superdiffusive regime. Moreover, in
a normal diffusive regime the maximum of the probability density function
(pdf), $f_{max}$, decreases as $f_{max}(t)\propto t^{-0.5}$, while in
superdiffusive and subdiffusive ones it decays as $f_{max}(t)\propto t^{-\gamma>0.5}$
and $f_{max}(t)\propto t^{-\gamma<0.5}$, respectively. Similar decays
exist also for the so-called Full Width at Half Maximum (FWHM) of
the pdf, namely $\textrm{FWHM}\propto t^{\gamma>0.5}$ and $\textrm{FWHM}\propto t^{\gamma<0.5}$
for super- and subdiffusive regimes. 

\section{Time and space generalized diffusion on graphs/networks}

Here we define a time and space generalized diffusion equation on
graphs/networks in the following way. Let $D_{t}^{\alpha}$ be the
Caputo fractional derivative and
let $\tilde{L}(s)$ be the Mellin-transformed $d$-path Laplacian
operator on $G$. Then, the generalized diffusion equation (GDE) is
defined as
\begin{align}
D_{t}^{\alpha}f(t)=-\tilde{L}(s)f(t), & f\left(0\right)=f_{0},\label{eq:GDE}
\end{align}
where

\begin{align}
D_{t}^{\alpha}f(t)=\frac{1}{\Gamma(1-\alpha)}\int_{0}^{t}\frac{f'(\tau)}{(t-\tau)^{\alpha}}d\tau & ,
\end{align}
$f'(\tau)$ denotes the
usual derivative. Here, $0<\alpha\leq1$ and $0<s<\infty$. 

Obviously, when $s\to\infty$, we have $D_{t}^{\alpha}f(t)=-Lf(t),$
where $L$ is the standard graph Laplacian. This equation accounts
for a time-fractional process only, without any spatial long-range
jumps. On the other hand, when $\alpha=1$ we get $\tfrac{df\left(t\right)}{dt}=-\tilde{L}(s)f(t),$
which accounts for long-range spatial jumps in the graph as studied
previously in \cite{estrada_path_2017}.

Our first result is the general expression for the solution
of the GDE defined before. We state this result in the following.
\begin{thm}
The solution of the GDE (\ref{eq:GDE}) is given by 
\begin{align}
f(t)=E_{\alpha}(-\tilde{L}(s)t^{\alpha})f_{0},\label{matrix_solution}
\end{align}where $E_{\alpha}(...)$ is the Mittag-Leffler  function of
the corresponding matrix:
\begin{align}
    E_{\alpha}(-\tilde{L}(s)t^{\alpha})=\sum_{j=0}^\infty \frac{(-\tilde{L}(s)t^{\alpha})^j}{\Gamma(\alpha j +1)}.
\end{align}
\end{thm}

\begin{proof}
Let $\tilde{L}(s)=U\Lambda U\adj$, where $\Lambda=diag(\lambda_{1},...,\lambda_{n})$
and $U$ is a unitary matrix of eigenvectors of $\tilde{L}(s)$. The dagger symbol denotes the conjugate transpose. Thus:
\begin{align}
D_{t}^{\alpha}f(t)=-U\Lambda U\adj f(t).
\end{align}
Defining $y(t)=U\adj f(t)$, we obtain a set of decoupled fractional
differential equations: 
\begin{align}
 & D_{t}^{\alpha}y_{i}(t)=-\lambda_{i}y_{i}(t), \\
 & y_{i}(0)=(U\adj f_{0})_{i}\eqqcolon y_{0i}.
\end{align}
Let us use the Laplace transform of this fractional differential equation,
and using the properties $\mc L\{D_{t}^{\alpha}y\}(u)=u^{\alpha}\mc L\{y\}(u)-u^{\alpha-1}y(0)$
as well as $\mc L\{E_{\alpha}(-at^{\alpha})\}(u)=\frac{u^{\alpha-1}}{u^{\alpha}+a}$,
we find the solution: 
\begin{align}
y_{i}(t)=y_{0i}E_{\alpha}(-\lambda_{i}t^{\alpha}) & .
\end{align}
Finally, we undo the change of basis $y(t)=U\adj f(t)$ to obtain:
\begin{align}
f(t)=E_{\alpha}(-\tilde{L}(s)t^{\alpha})f_{0} & ,  \label{solution_time_independent}
\end{align}
where the definition of a matrix function $g$ acting on a diagonalizable
matrix $A=U\Lambda U\adj$ is: $g(A)=Ug(\Lambda)U\adj$, and $g(\Lambda):=\textnormal{diag}(g(\lambda_{i})).$
\end{proof}

\subsection{Analysis of different diffusive regimes}

We now focus on understanding how the GDE can give rise to different
diffusive regimes in a given graph. In particular, we will focus here
only in the one-dimensional case. For that, we will consider an infinite
path graph, $P_{\infty}$, which corresponds to the one-dimensional
case. Anomalous diffusion in one-dimension has received much attention
in the literature \cite{1D_0,1D_1,1D_2,1D_3,1D_4} due to its practical
relevance.

Let us start by defining the Fourier transform operator $\mc F$ and
its inverse $\mc F^{-1}$, respectively: 
\begin{align}
 & \mc F\{f\}(k)=\onov{\sqrt{2\pi}}\sum_{n\in\Z}f_{n}e^{ink},\\
 & (\mc F^{-1}\{g\})_{m}=\onov{\sqrt{2\pi}}\int_{-\pi}^{\pi}e^{-imk}g(k)dk.
\end{align}

Let us first obtain the solution of the GDE for the $P_{\infty}$
graph.
\begin{lem}
The fundamental solution of the GDE for the infinite path graph $P_{\infty}$
and initial condition $f_{n}(0)=f_0=\delta_{n0}$, can be expressed as:
\begin{align}
f_{m}(t)=\onov{2\pi}\int_{-\pi}^{\pi}dke^{-ikm}E_{\alpha}(-t^{\alpha}l_{s}(k)),
\end{align}
with
\begin{align}
    l_{s}(k)=2\zeta(s)-\Li_{s}(e^{ik})-\Li_{s}(e^{-ik}), 
\end{align}
where $\zeta(s)=\sum_{k=1}^{\infty}\frac{1}{k^{s}}$ is the Riemann zeta function and $\Li_{s}(x)=\sum_{n=1}^{\infty}\frac{x^{n}}{n^{s}}$
is the polylogarithm. 
\end{lem}

\begin{proof}
The Fourier transform of the Mellin-transformed $d$-path Laplacian
operator $\tilde{L}(s)$ is \cite{estrada_path_2017}: 
\begin{align}
\mc F\{\tilde{L}(s)f_{n}\}=l_{s}(k)\mc F\{f_{n}\},
\end{align}
therefore: 
\begin{align}
\mc F\{f(t)\} = \mc F\{E_{\alpha}(-\tilde{L}(s)t^{\alpha})f_{0}\}=E_{\alpha}(-l_{s}(k)t^{\alpha})\mc F\{f_{0}\} & .
\end{align}
Using the inverse Fourier transform and substituting the initial condition $f_{n}(0)=(f_{0})_{n}=\delta_{n0}$, we finally arrive at the expression: 
\begin{align}
f_{m}(t)=\onov{2\pi}\int_{-\pi}^{\pi}dke^{-ikm}E_{\alpha}(-t^{\alpha}l_{s}(k)).
\end{align}
\end{proof}
The time scaling of this expression is unclear, due to the polylogarithms inside the Mittag-Leffler function. However, to determine
the diffusive regime that the system exhibits, it is sufficient to
characterize the behavior of the solution in the limit of long times.
Thus, we perform an asymptotic approximation, which is accounted for
in the following result.
\begin{lem} \label{lem:4}
Let $\beta>0$ and let $l:[-\pi,\pi]\to\R$ be a continuous function
satisfying: 
\begin{align}
 & l(k)>0\quad for\ \ k\in[-\pi,\pi]\backslash0\\
 & l(k)\sim c|k|^{\beta}\quad as\ \ k\to0\label{hyp1}
\end{align}
with some $c>0$. $A(k)\sim B(k)$ denotes that $\lim_{k\to0}\frac{A(k)}{B(k)}=1$.
Then: 
\begin{align}
\frac{1}{2\pi}\int_{-\pi}^{\pi}e^{-ikm}E_{\alpha}(-t^{\alpha}l(k))dk\to\frac{1}{2\pi}\int_{-\pi}^{\pi}e^{-ikm}E_{\alpha}(-ct^{\alpha}|k|^{\beta})dk=\mc F^{-1}\left\{ \onov{\sqrt{2\pi}}E_{\alpha}(-ct^{\alpha}|k|^{\beta})\right\} \label{lemma2}
\end{align}
as $t\to\infty$.
\end{lem}

\begin{proof}
Let $\epsilon>0$ and let us decompose the Fourier transform integral into:
\begin{align}
    \onov{2\pi}\int_{-\pi}^\pi e^{-ikm}E_\alpha(-t^\alpha l(k))dk=\onov{2\pi} \int_{-\epsilon}^\epsilon e^{-ikm}E_\alpha(-t^\alpha l(k))dk +  \onov{2\pi} \int_{[-\pi,-\epsilon]\cup [\epsilon,\pi]} e^{-ikm}E_\alpha(-t^\alpha l(k))dk.  \label{integrals_aux}
\end{align}
Let us take the limit $\epsilon\to 0$, so that the first integral becomes:
\begin{align}
    \int_{-\epsilon}^\epsilon e^{-ikm}E_\alpha(-t^\alpha l(k))dk\to \int_{-\epsilon}^\epsilon e^{-ikm}E_\alpha(-t^\alpha c|k|^\beta)dk.
\end{align}
The second integral from \eqref{integrals_aux} is negligible:
\begin{align}
    \Bigg |\int_{[-\pi,-\epsilon]\cup [\epsilon,\pi]} e^{-ikm}E_\alpha(-t^\alpha l(k))dk\Bigg |\leq \int_{[-\pi,-\epsilon]\cup [\epsilon,\pi]} E_\alpha(-t^\alpha l(k))dk=2 \int_\epsilon^\pi E_\alpha(-t^\alpha l(k))dk\to 0  \quad \textrm{as} \quad t\to\infty.
\end{align}
In particular, for a given $\epsilon$, the integral is negligible whenever $t\gg l(\epsilon)^{-1/\alpha}$. \\
For identical reasons, the integral
\begin{align}
    \int_{[-\pi,-\epsilon]\cup [\epsilon,\pi]} e^{-ikm}E_\alpha(-t^\alpha c |k|^\beta)dk
\end{align}
is also negligible when $t\to\infty$. Thus, we can express the Fourier transform integral as:
\begin{align}
    \onov{2\pi} \int_{-\pi}^\pi e^{-ikm}E_\alpha(-t^\alpha l(k))dk\to \onov{2\pi} \int_{-\pi}^\pi e^{-ikm}E_\alpha(-t^\alpha c |k|^\beta)dk=\mc F^{-1}\left \{\onov{\sqrt{2\pi}}  E_\alpha(-t^\alpha c |k|^\beta)  \right \}.
\end{align}
\end{proof}
\begin{cor}
The solution of the GDE, when $t\to\infty$, is given by: 
\begin{align}
f_{m}(t)=\frac{1}{2\pi}\int_{-\pi}^{\pi}e^{-ikm}E_{\alpha}(-ct^{\alpha}|k|^{\beta})dk=\mc F^{-1}\left\{ \onov{\sqrt{2\pi}}E_{\alpha}(-ct^{\alpha}|k|^{\beta})\right\}  & ,\label{solution_GDE}
\end{align}
with
\begin{align}
\beta=\begin{cases}
s-1\quad\textrm{if}\quad1<s<3,\\
2\quad\quad\ \ \textrm{if}\quad s>3.
\end{cases}\label{beta}
\end{align}
\end{cor}

\begin{proof}
According to Theorem 6.5 of \cite{estrada_path_2017}, the function $l_{s}(k)=2\zeta(s)-\Li_{s}(e^{ik})-\Li_{s}(e^{-ik})$ has the following asymptotics as $k\to0$:
\begin{align}
l_{s}(k)\sim\begin{cases}
-\frac{\pi}{\Gamma(s)\cos\left(\frac{s\pi}{2}\right)}|k|^{s-1}\quad\textrm{if}\quad1<s<3,\\
\zeta(s-2)k^{2}\quad\quad\quad\quad\ \ \textrm{if}\quad s>3,
\end{cases}
\end{align}
Moreover, $l_{s}(k)$ is continuous and fulfills $l_{s}(k)>0\ \forall k\neq0,\ l_{s}(0)=0$ (\cite{estrada_path_2017}, Lemma 5.1).
Thus, $l_s(k)$ fulfills the hypotheses of Lemma \ref{lem:4}, so its asymptotic approximation is given by eq. \eqref{lemma2}.
\end{proof}

Equation \eqref{solution_GDE} fully determines the solution of the
GDE for long times. Unfortunately, the Fourier transform cannot be
solved in terms of elementary functions. Nevertheless, the time scaling can be obtained with the following Theorem. 
\begin{thm}  \label{thm:6}
The solution of the GDE $f(t)$ has the asymptotic scaling with time $f(t)\propto t^{-\frac{\alpha}{\beta}}$, as $t\to\infty$.
\end{thm}

\begin{proof}
Let us exploit the parity of the integrand in eq. \eqref{solution_GDE} and expand the cosine into a power series, we obtain:
\begin{align}
    f_{m}(t)&=\onov{\pi}\int^\pi_{0} dk \cos(km) E_\alpha(-c k^\beta  t^\alpha)  = 
    \sum_{j}\onov{\pi}\int^\pi_{0} dk (-1)^j \frac{(km)^{2j}}{(2j)!} E_\alpha(-c k^\beta  t^\alpha)  .
\end{align}
We perform the change of variables $u=c k^\beta  t^\alpha$ and split the integration interval into two:
\begin{align}  \label{integrals}
    \int^\pi_{0}dk  k^{2j} E_\alpha(-c k^\beta  t^\alpha) &= \frac{1}{\beta (c t^\alpha)^{\frac{2j+1}{\beta }}}\left(\int^{\infty}_{0} du u^{\frac{2j+1-\beta}{\beta }} E_\alpha(-u)   -  \int^{\infty}_{\pi^\beta  c  t^\alpha}du  u^{\frac{2j+1-\beta}{\beta }} E_\alpha(-u)  \right).
\end{align}
The second integral is bounded by the constant $e^{-\frac{\pi^{\beta}ct^{\alpha}}{2}}$, as we will show now. For sufficiently big $u$, $E_{\alpha}(- u)<e^{- u}$
and $ u^{\frac{2j+1-\beta}{\beta}}<e^{\frac{ u}{2}}$,
so: 
\begin{align}
\int_{\pi^{\beta}ct^{\alpha}}^{\infty}E_{\alpha}(- u) u^{\frac{2j+1-\beta}{\beta}}d u<\int_{\pi^{\beta}ct^{\alpha}}^{\infty}e^{- u}e^{\frac{ u}{2}}d u=e^{-\frac{\pi^{\beta}ct^{\alpha}}{2}} & .
\end{align}
In the limit $t\to\infty$, $e^{-\frac{\pi^{\beta}ct^{\alpha}}{2}}$ approaches zero and thus the second integral from eq. \eqref{integrals} is negligible; therefore:
\begin{align}
    f_{m}(t)&=
    \sum_{j}\frac{(-1)^j}{\pi}   \frac{m^{2j}}{(2j)!} \frac{1}{\beta (c t^\alpha)^{\frac{2j+1}{\beta }}} I_j(\alpha,\beta) = \onov{\pi \beta}  \left(  \frac{I_0(\alpha,\beta)}{ c^{\frac{1 }{\beta }}} t^{-\frac{\alpha}{\beta}} -
    \frac{m^{2}}{2} \frac{I_1(\alpha,\beta)}{ c^{\frac{3}{\beta }}}t^{-\frac{3\alpha}{\beta}}  + ... \right)  ,  \label{solution_FT}
\end{align}
where $I_j(\alpha,\beta)=\int^{\infty}_{0} du u^{\frac{2j+1-\beta}{\beta }} E_\alpha(-u) $.  \\
For large times and small $m$ (i.e., in the central region), the dominant term in the sum is the one with the smallest absolute value of the exponent of $t$; i.e., the term $j=0$. Hence, the asymptotic scaling with time is
\begin{align}
    f_{m}(t)\propto t^{-\frac{\alpha}{\beta}}.
\end{align}
\end{proof}
Additionally, let us find the scaling of two relevant observables: the height of the maximum of the pdf and the Full Width at Half Maximum (FWHM).
\begin{lem}
The maximum of the pdf, $f_{max}$, and FWHM fulfill the following asymptotic scaling laws in time: \\
i) $f_{max}(t) \propto t^{-\frac{\alpha}{\beta}}$, \\
ii) $\textrm{FWHM}(t) \propto  t^{\frac{\alpha}{\beta}}$
\end{lem}
\begin{proof}
 The proof of (i) is straightforward using eq. \eqref{solution_FT} for $m=0$:
 \begin{align}
     f_{max}(t) = \onov{\pi \beta}    \frac{I_0(\alpha,\beta)}{ c^{\frac{1 }{\beta }}} t^{-\frac{\alpha}{\beta}} \propto t^{-\frac{\alpha}{\beta}}.
 \end{align}

To prove (ii), we make the rescaling $m\to mt^{\frac{\alpha}{\beta}}$
in eq. \eqref{solution_GDE}. The function $f(m,t)$ becomes: 
\begin{align}
f(mt^{\frac{\alpha}{\beta}},t)=\onov{2\pi}\int_{-\pi}^{\pi}e^{-ikt^{\frac{\alpha}{\beta}}m}E_{\alpha}(-ct^{\alpha}|k|^{\beta})dk=t^{-\frac{\alpha}{\beta}}\onov{2\pi}\int_{-\pi}^{\pi}e^{-i\tilde{k}m}E_{\alpha}(-c|\tilde{k}|^{\beta})d\tilde k=t^{-\frac{\alpha}{\beta}}f(m,1),  \label{scaling_relation}
\end{align}
where $\tilde{k}=t^{\frac{\alpha}{\beta}}k$. 
Let us apply this equation to find the scaling of the FWHM. Let $\xi(t)$ be the position at which the pdf reaches its Half Maximum at time $t$ (i.e., the "Half Width at Half Maximum"), so that $\textrm{FWHM}(t)=2\xi(t)$. The FWHM can be determined through the equation $f(\xi(t),t)=\frac{f_{max}(t)}{2}$. Using the time scaling of the maximum obtained in (i), we can express this as:
\begin{align}
    f(\xi(t),t)=\frac{f_{max}(1)t^{-\frac{\alpha}{\beta}}}{2}=f(\xi(1),1)t^{-\frac{\alpha}{\beta}}.
\end{align}
Comparing this equation with the scaling relation \eqref{scaling_relation}, it follows that $\xi(t)=\xi(1)t^{\frac{\alpha}{\beta}}$, and thus:
\begin{align}
    \textrm{FWHM} \propto t^{\frac{\alpha}{\beta}}.
\end{align}
\end{proof}
\textbf{Remark:} The MSD for processes with infinite variance of the step size distribution diverges \cite{jespersen_levy_1999, dybiec_levy_2017}, unless the step size distribution and the waiting time distribution are correlated \cite{klafter_stochastic_1987}. This is a well-known limitation of Levy flights and other stochastic processes that lead to superdiffusion \cite{dybiec_levy_2017}. Proposals of pseudo mean square displacements that remain finite for superdiffusive processes have been made \cite{jespersen_levy_1999, metzler_random_2000}, but their analytical calculation is far from trivial. \\

Let us now recall that for normal diffusion, the corresponding scaling
laws are: i) $f(t)\propto t^{-\frac{1}{2}}$, ii) $f_{max}(t)\propto t^{-\frac{1}{2}}$,and iii) $\textrm{FWHM}(t)\propto t^{\frac{1}{2}}$. Let us focus here
on the case $1<s<3$, where $\beta=s-1$. We find that the three scaling
laws predict the same location of the three possible diffusive regimes: 

Standard diffusion: $\alpha=\frac{s-1}{2}$. 

Superdiffusion: $\alpha>\frac{s-1}{2}$. 

Subdiffusion: $\alpha<\frac{s-1}{2}$. 

\begin{figure}
\centering{}\includegraphics[scale=0.75]{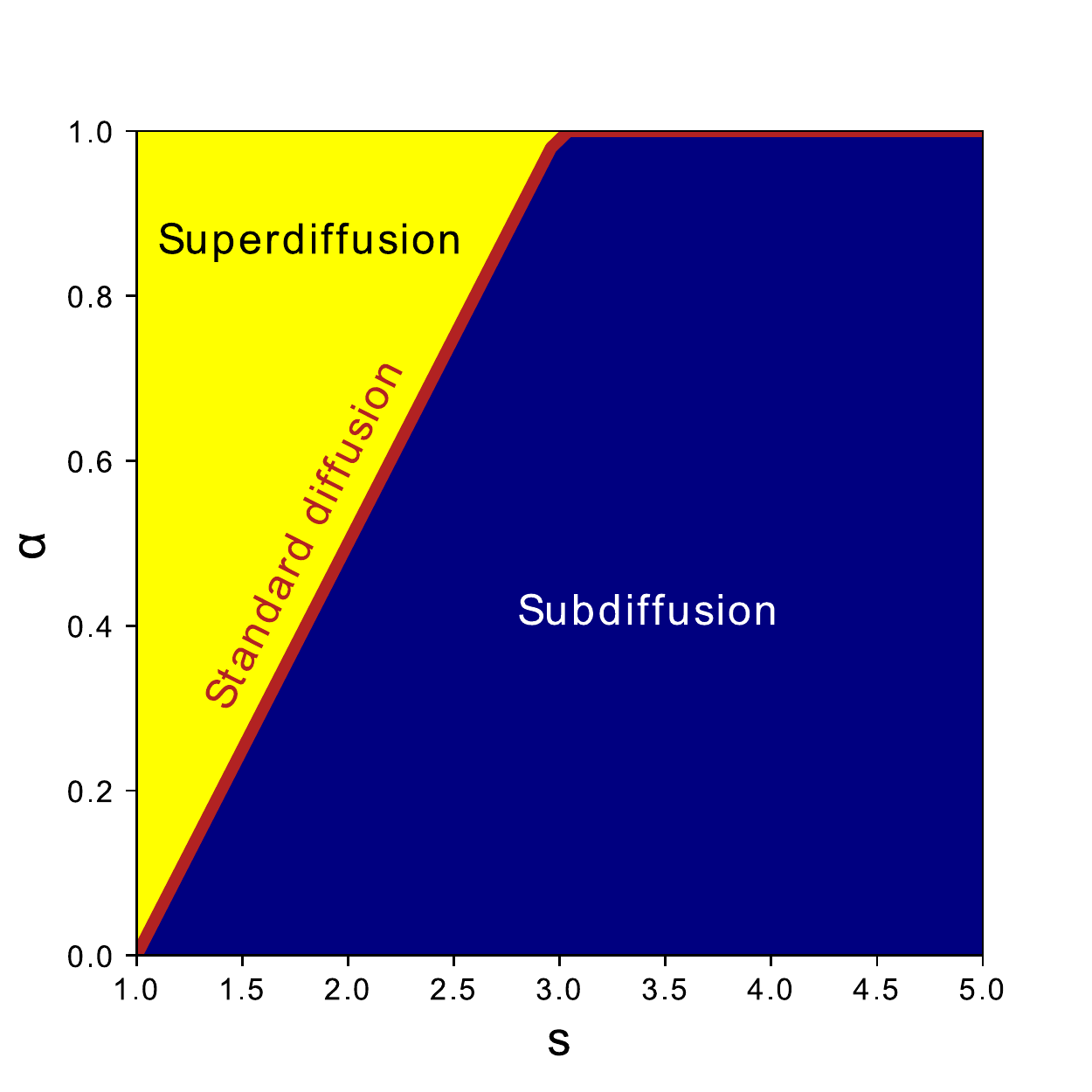}
\caption{Phase diagram with the three possible diffusive regimes (superdiffusion,
subdiffusion and standard diffusion) as a function of the parameters
$\alpha$ and $s$.}
\label{fig:phase_diag} 
\end{figure}

For $s>3$, $\beta=2$, which implies that subdiffusion happens for
any $\alpha\in(0,1)$, and standard diffusion for $\alpha=1$. A phase
diagram with the different diffusive regimes can be found in Figure
\ref{fig:phase_diag}. 

\section{Time-dependent generalized diffusive processes}

Let us consider here a system in which the different diffusive regimes
alternate with time. That is, a system in which the diffusive particle
behaves subdiffusively at a given time, then alternates with normal diffusion and superdiffusion as the time goes on. These processes can alternate cyclically as we will see later on this work. 

\subsection{Stepwise temporal dependence}

The simplest way of achieving this is to let the parameter $s$ depend
on time. This idea was previously explored by Allen-Perkins et al. \cite{allen-perkins_approach_2019}
for the case of the $d$-path Laplacian operator. Let us define:
\begin{align}
 & \tilde{L}(s(t))=\sum L_{d}d^{-s(t)},\label{tdep_L_1}\\
 & s(t)=\begin{cases}
s_{1}\wif t<t_{sw},\\
s_{2}\wif t>t_{sw}.
\end{cases}\label{tdep_L_2}
\end{align}
 We have the following generalized diffusion equation: 
\begin{align}
 & D_{t}^{\alpha}{f\left(t\right)}=-\tilde{L}(s(t))f\left(t\right),
  f(0)=f_{0}.
\end{align}
For a step-like $s$ like eq. \eqref{tdep_L_2}, one can split the
system into two time-independent equations: 
\begin{align}
D_{t}^{\alpha}f_A(t)=-\tilde{L}(s_{1})f_A(t)\wif t<t_{sw} & ,\\
D_{t}^{\alpha}f_B(t)=-\tilde{L}(s_{2})f_B(t)\wif t>t_{sw} & ,
\end{align}
with initial condition $f_A(0)=f_{0}$. Then, expressing $\tilde{L}(s_{q})$
$(q\in \{A, B\})$ as $\tilde{L}(s_{q})=U_{q}\Lambda_{q}U_{q}\adj$, where
$U_{q}$ are unitary and $\Lambda_{q}=\textnormal{diag}((\lambda_{q})_{i})$,
and defining $y_{q}=U_{q}\adj f_q$, we get a set of decoupled
differential equations. The solution of each one is:
\begin{align}
(y_{q})_{i}(t)=E_{\alpha}(-(\lambda_{q})_{i}t^{\alpha})(\tilde{C}_{q})_{i} & ,
\end{align}
where $\tilde{C}_{q}$ are integration constants. $\tilde C_A$ can be obtained using the initial condition $f_A(0)=f_{0}$ and $\tilde C_B$ is found imposing the continuity of the function at $t=t_{sw}$: $f_{A}(t_{sw})=f_{B}(t_{sw})$. Undoing the change of basis induced by $U_q\adj$, we obtain the following solution in terms of matrix functions:
\begin{align}
f(t)=\begin{cases}
E_{\alpha}(-\tilde L(s_{1})t^{\alpha})f_{0} & \wif{t\leq t_{sw}},\\
E_{\alpha}(-\tilde L(s_{2})t^{\alpha})(E_{\alpha}(-\tilde L(s_{2})t_{sw}^{\alpha}))^{-1}E_{\alpha}(-\tilde L(s_{1})t_{sw}^{\alpha})f_{0} & \wif{t>t_{sw}}.
\end{cases}\label{time_dep_s_exp_sol}
\end{align}
From now on, we assume that $t>t_{sw}$. We express eq.
\eqref{time_dep_s_exp_sol} as: 
\begin{align}
f(t)=E_{\alpha}(-\tilde L(s_{2})t^{\alpha})f_{1},
\end{align}
where $f_{1}(t_{sw};s_{1},s_{2})=E_{\alpha}(-\tilde L(s_{2})t_{sw}^{\alpha})^{-1}E_{\alpha}(-\tilde L(s_{1})t_{sw}^{\alpha})f_{0}$. 
The introduction of $f_1$ reduces the problem to a time-independent expression, analogous to eq. \eqref{solution_time_independent}, but with a different initial condition. The time-evolved system
is then 
\begin{align}
f_{m}(t)=\onov{2\pi}\sum_{n}\int_{-\pi}^{\pi}dke^{-ik(m-n)}E_{\alpha}(-t^{\alpha}l_{s_{2}}(k))(f_{1})_{n}\to \mc F^{-1}\left\{  E_{\alpha}(-c_{2}t^{\alpha}|k|^{\beta_{2}})\mc F\{f_{1}\}(k)\right\}  &   \label{time_evolved_stepwise}
\end{align} 
as $t\to\infty$, where we have used the asymptotic equivalence $l_{s_2}(k)\sim c_2 |k|^{\beta_2}$ as $k\to 0$ and Lemma \ref{lem:4}. 
The time scaling of the pdf can be calculated using the procedure from Theorem \ref{thm:6}, obtaining: 
\begin{align}
    f_{m}(t)&=
    \sum_{j,n}\frac{(-1)^j}{\pi}   \frac{(m-n)^{2j}}{(2j)!} \frac{1}{\beta_2 (c_2 t^\alpha)^{\frac{2j+1}{\beta_2 }}} I_j(\alpha,\beta_2) (f_1)_n,  \label{tdep_solution}
\end{align}
which confirms that the dominant time scaling is still $t^{-\frac{\alpha}{\beta_2}}$. In other words, the phase diagram from the previous section remains unchanged: the system behaves asymptotically as if the Laplacian $L(s_1)$ had never affected the dynamics.

\subsection{Periodic temporal alternancy}

Consider the following time-dependent parameter $s$, with period $T$:
\begin{align}
s(t)=
    \begin{cases}
        s_1 \wif nT\leq t<(n+\frac{1}{2})T, \\
        s_2 \wif (n+\frac{1}{2})T\leq t<(n+1)T,
    \end{cases} 
\end{align}
where $n\in\N \cup \{0\}$. In order to obtain the full solution of the problem, let us define $f^{(n)}(t)$ as the restriction of $f(t)$ to $t\in[nT,(n+1)T)$: $f^{(n)}(t)=f(t)|_{[nT,(n+1)T)}$. To solve the problem, it suffices to find an expression for $f^{(n)}(t)$ for arbitrary $n$.
\begin{thm}
The solution of the GDE with a periodic parameter $s(t)$ and initial condition $f(0)=g_0$ is given by:
\begin{align}
    f^{(n)}(t)=  
    \begin{cases}
    E_{\alpha}(-L(s_1) t^\alpha)g_n & \wif{nT\leq t<(n+\frac{1}{2})T}\\
     E_{\alpha}(-L(s_2) t^\alpha) 
     (E_{\alpha}(-L(s_2) \left(nT+\frac{T}{2}\right)^\alpha))^{-1} E_{\alpha}(-L(s_1) \left(nT+\frac{T}{2}\right)^\alpha)g_n  & \wif{(n+\frac{1}{2})T\leq t<(n+1)T}, \label{solution_periodic_s}
     \end{cases}
\end{align}
where:
\begin{align}
    g_n:=&E_{\alpha}(-L(s_1) (nT)^\alpha)^{-1}                E_{\alpha}(-L(s_2) (nT)^\alpha)
     (E_{\alpha}(-L(s_2) \left(nT-\frac{T}{2}\right)^\alpha))^{-1} E_{\alpha}(-L(s_1) \left(nT-\frac{T}{2}\right)^\alpha)g_{n-1}= \nonumber \\  
     =& \prod_{j=1}^n E_{\alpha}(-L(s_1) (jT)^\alpha)^{-1}                E_{\alpha}(-L(s_2) (jT)^\alpha)
     (E_{\alpha}(-L(s_2) \left(jT-\frac{T}{2}\right)^\alpha))^{-1} E_{\alpha}(-L(s_1) \left(jT-\frac{T}{2}\right)^\alpha) g_0.
\end{align}  
\end{thm}
\begin{proof}
 $f^{(0)}(t)$ is given by eq. \eqref{time_dep_s_exp_sol}, for $t_{sw}=\frac{T}{2}$. The remaining $f^{(n)}(t)$ can be obtained by induction. Suppose $f^{(n)}(t)$ is given by eq. \eqref{solution_periodic_s}. The solution to the GDE in the interval $[(n+1)T,(n+2)T)$ is:
     \begin{align}
     f^{(n+1)}(t)=
         \begin{cases}
            E_{\alpha}(-L(s_1) t^\alpha)C_1 & \wif{(n+1)T\leq t<((n+\frac{3}{2})T},\\
            E_{\alpha}(-L(s_2) t^\alpha) C_2  & \wif{(n+\frac{3}{2})T\leq t<(n+2)T}.
     \end{cases}
     \end{align}
where $C_1, C_2$  are integration constants. \\
Let us impose the continuity of the function at $t=(n+1)T$ and $t=(n+\frac{3}{2})T$ to obtain the values of the integration constants:
\begin{align}
    C_1=&E_\alpha(-L(s_1)((n+1)T)^\alpha)^{-1}   E_{\alpha}(-L(s_2) ((n+1)T)^\alpha) \times \nonumber \\ & \times (E_{\alpha}(-L(s_2) \left(nT+\frac{T}{2}\right)^\alpha))^{-1}  E_{\alpha}(-L(s_1) \left(nT+\frac{T}{2}\right)^\alpha)g_n =g_{n+1}, \\
    C_2=&(E_\alpha(-L(s_2)((n+\frac{3}{2})T)^\alpha)^{-1}E_{\alpha}(-L(s_1) ((n+\frac{3}{2})T)^\alpha)g_{n+1}.
\end{align}
This corresponds to eq. \eqref{solution_periodic_s} for $n\to n+1$. Thus, by induction, expression \eqref{solution_periodic_s} holds for any $n$.
\end{proof}
The pdf given by \eqref{tdep_solution} is still valid, with $f_1=g_n \ \textrm{if} \ {nT\leq t<((n+\frac{1}{2})T}$ and $f_1=(E_{\alpha}(-L(s_2) \left(nT+\frac{T}{2}\right)^\alpha))^{-1} \times $\\
$\times E_{\alpha}(-L(s_1) \left(nT+\frac{T}{2}\right)^\alpha)g_n   \ \textrm{if} \ {(n+\frac{1}{2})T\leq t<(n+1)T}$. This implies that the time scaling at time $t$ depends only on the parameter $s$ of the system at that time. Consequently, the phase diagram of fig. \ref{fig:phase_diag} is valid, adapting $s$ to the Mellin transform parameter that acts at the desired time.

\section{Generalized diffusion on a linear chain}

\begin{figure}
\includegraphics[width=1\textwidth]{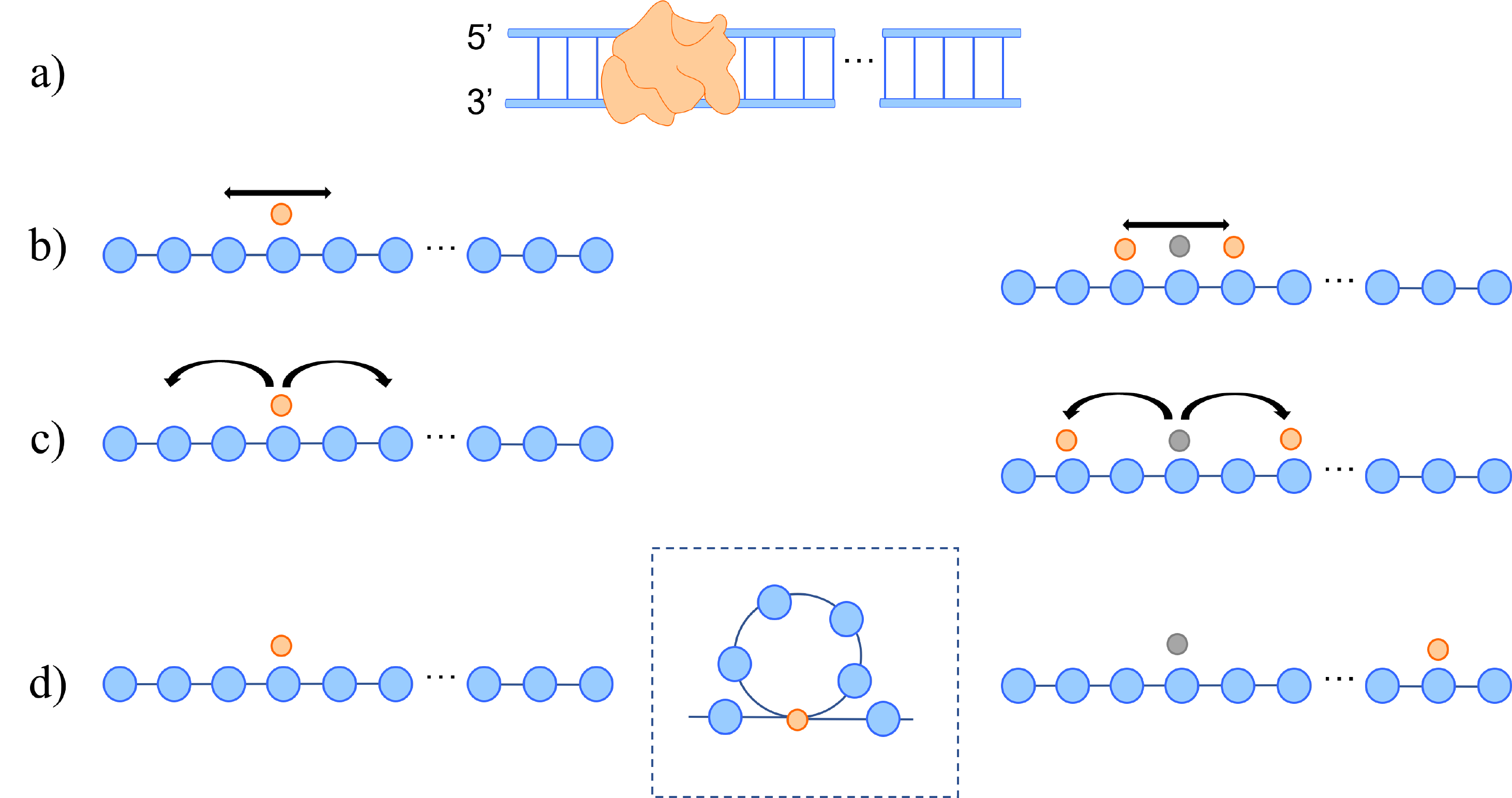}

\caption{(a) Representation of DNA chains and the transcription factor (TF).
In the following panels DNA is represented as a chain graph where
every pair of bases is represented by a node and two pairs of consecutive
nodes are connected by an edge. (b-d) Different processes by which
TF inspect and repair DNA: (b) sliding, (c) hopping and (d) intersegmental
transfer.}

\label{DNA}
\end{figure}

We consider here the diffusion of a particle along a linear chain
or path graph $P_{n}$. The motivation of these simulations is provided
by the diffusion of a protein along a DNA chain. We consider the double
chains of DNA (see Fig. \ref{DNA} a) as the path graph $P_{n}$.
In doing so, we consider a pair of DNA bases, one from the 5' chain
and another from the 3' chain, as a node of the path graph. The protein
is then modeled as a particle diffusing on the linear chain. We consider
here that the protein can diffuse by a combination of the following
mechanisms:\\
(i) one-dimensional diffusion or sliding, involving a
random walk along the DNA without dissociation as illustrated in Fig.
\ref{DNA} b; \\
(ii) jumping, where a protein moves over longer distances
via dissociation and then rebinding at a distal location (see Fig.
\ref{DNA} c); \\
(iii) intersegmental transfer, involving movement from
one site to another via a looped intermediate (Fig. \ref{DNA} d). \\
The first process gives rise to normal diffusion, but we consider
that at certain DNA regions the protein makes a longer exploration/repair
which acts as a trap producing subdiffusive motion, which can be modeled
by varying the exponent $\alpha$ of the Caputo fractional derivative.
The mechanisms (ii) and (iii) clearly represent long-jumps controlled
by the Mellin exponent in the transformed $d$-path Laplacian. We
should remark that other approaches have been reported in the literature
to account for some aspects of the protein diffusion on DNA \cite{DNA_1,DNA_2,DNA_3}.

For the simulations we consider a path graph with $N=1001$ nodes.
We always take as initial condition $f_{n}(0)=\delta_{n,501}$, i.e.,
we locate all the diffusive particle at the center of the linear chain.
To obtain the probability density functions, we calculated eq. \eqref{matrix_solution}
using the algorithm from \cite{garrappa_computing_2018} for matrix
Mittag-Leffler functions. We then fitted the time evolution of the maximum of the pdf and FWHM to a power law $t^{\gamma}$, and compared the numerical exponent $\gamma$
with the corresponding theoretical prediction. We
have noticed that in the superdiffusive regime it is hard to precisely
quantify the MSD, since the asymptotic behavior is reached for large
times and erratic behaviors can appear in the transient regime (see
\cite{Alves_characterization_2016} for a detailed discussion). Moreover,
the FWHM method, while providing correct predictions, introduces a
higher degree of inaccuracy in the exponent $\gamma$. Because of
this, we use the method of the decay of the maximum to find the numerical
exponent $\gamma$. 

\begin{figure}
\centering{}\includegraphics[width=0.8\linewidth]{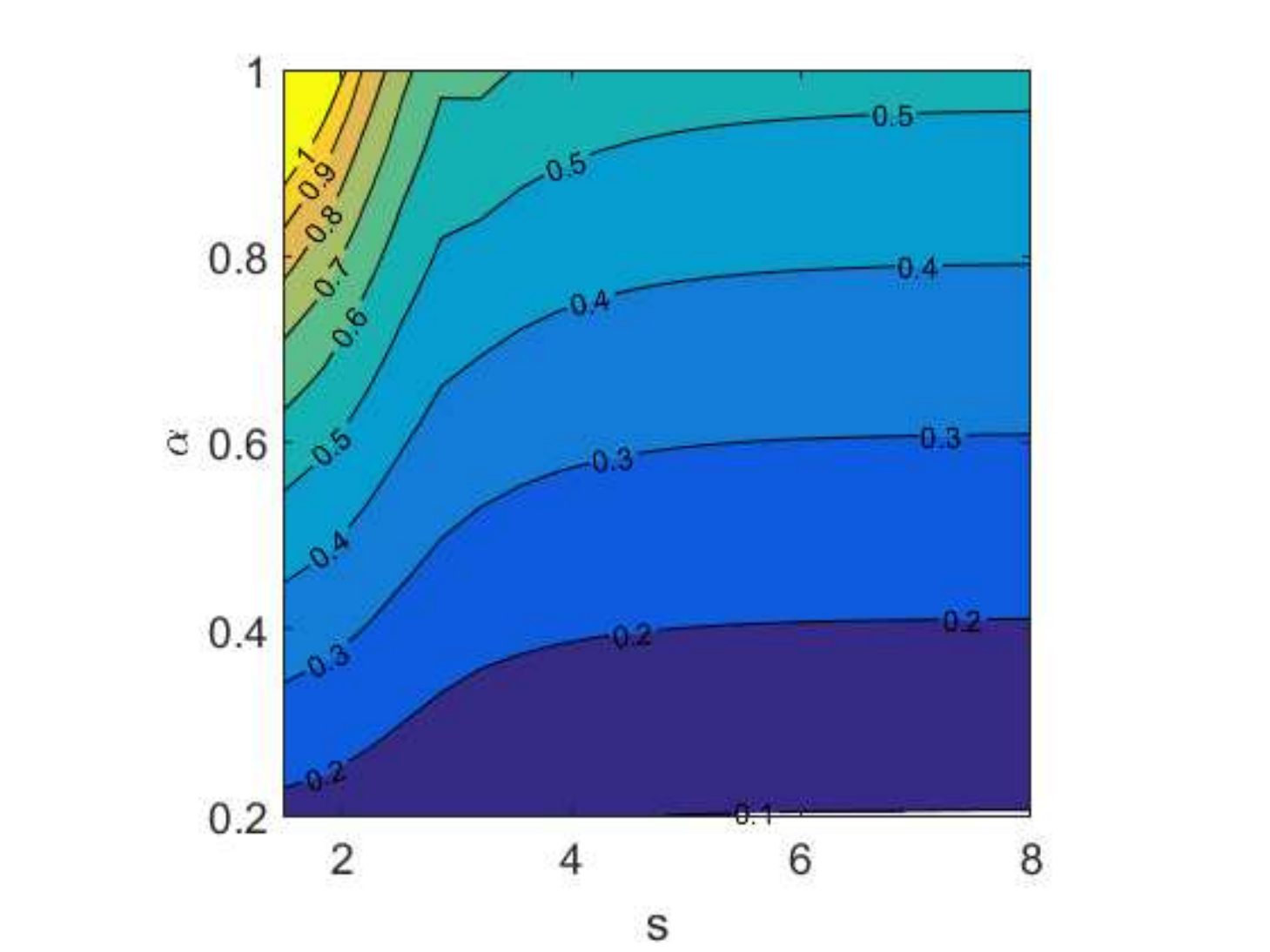}
\caption{Exponent $\gamma$ as a function of the control parameters $s$ and
$\alpha$. $\gamma>0.5$ corresponds to superdiffusion and $\gamma<0.5$,
to subdiffusion. The exponents $\gamma$ were obtained by fitting
the decay of the maximum to a power law. For each pair ($s$, $\alpha$),
the time range used to measure this decay was $t\in[0,10^{\frac{\beta}{\alpha}}]$, where $\beta$ is given by eq. \eqref{beta}.}
\label{fig:contour}
\end{figure}

Our first target is to obtain a contour plot indicating how the exponent
$\gamma$ defining the type of diffusive regime changes with the values
of the model parameters $\alpha$ and $s$. The results are illustrated
in Fig. \ref{fig:contour}. We recall that for $f_{\textnormal{max}}(t)\propto t^{-\gamma}$:

i) $\gamma<0.5$ represents subdiffusion;

ii) $\gamma=0.5$ represents normal diffusion;

iii) $\gamma>0.5$ represents superdiffusion.

Therefore, we can see the curve with the value $\gamma=0.5$ in the
Fig. \ref{fig:contour}, which indicates the normal diffusion regime.
When we move towards the right-upper corner of the plot, i.e., $s\rightarrow\infty$
and $\alpha\rightarrow1$, we are moving to a regime completely dominated
by normal diffusion, which is represented by the standard graph diffusion
equation. However, it is remarkable that even with strong long-jumps,
e.g., $s\approx2,$ we can still find regions of normal diffusion
if the temporal memory parameter $\alpha$ is relatively small, e.g.,
$\alpha\approx0.6$. Over this curve we are in a superdiffusive regime
and below it we are in the subdiffusive one. It is remarkable that
there is an abrupt change of behavior at $s=3$, which matches the
theoretical prediction that after this point the superdiffusive regime
no longer exists. It can be seen that for $s<3$ the contour lines
of $\gamma$ abruptly decay, while for $s>3$ they are almost parallel
to the $x$-axis. 

\begin{figure}
\subfloat[]{\begin{centering}
\includegraphics[width=0.33\textwidth]{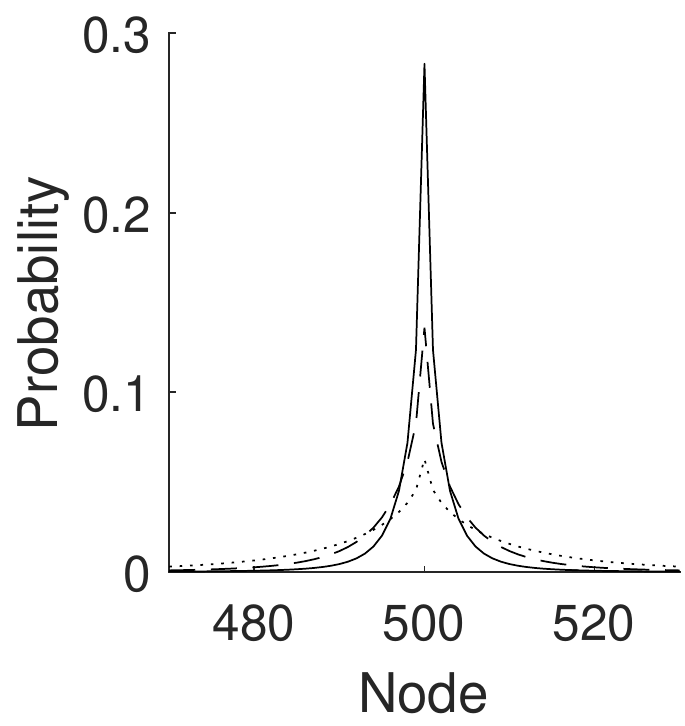}
\par\end{centering}
}\subfloat[]{\begin{centering}
\includegraphics[width=0.33\textwidth]{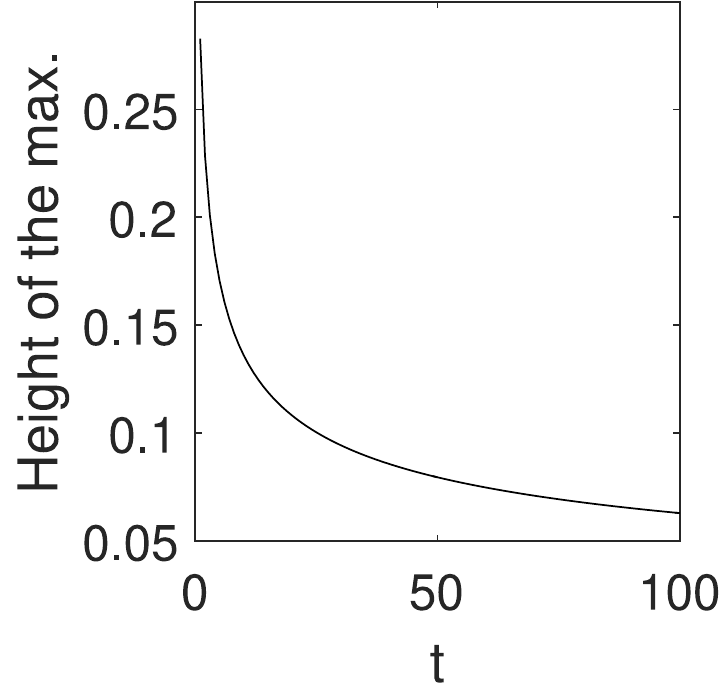}
\par\end{centering}
}\subfloat[]{\begin{centering}
\includegraphics[width=0.33\textwidth]{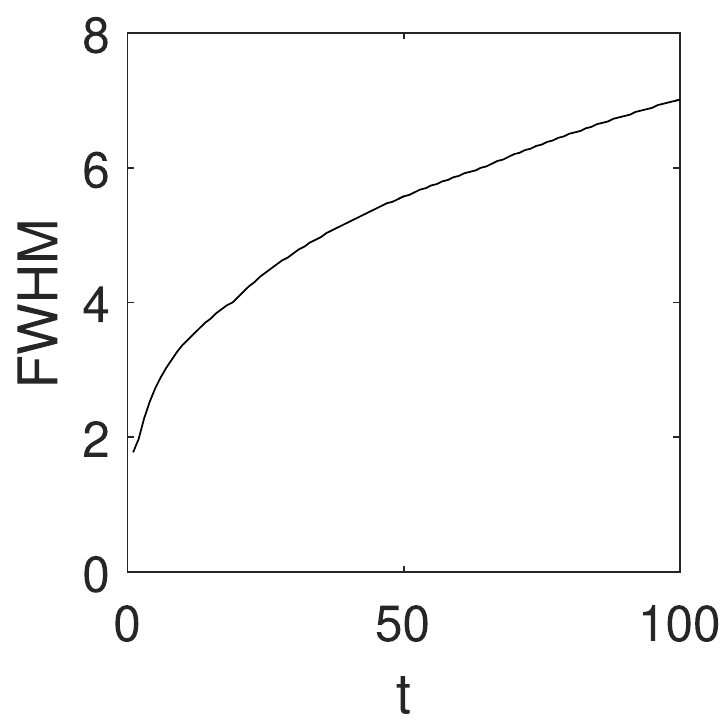}
\par\end{centering}
}

\caption{Illustration of the time evolution of a diffusive particle on a path
graph with $N=1001$ nodes, using $\alpha=0.5$, $s=2.5$. The pdf
is initially a delta distribution at the middle of the linear chain.
(a) Probability density function in the different nodes, for three different times: $t=1$ (solid line),$t=10$ (dashed line) and $t=100$ (dotted line). (b) Decay
of the height of the maximum with time. (c) Increase of the FWHM as
a function of time.}

\label{example}
\end{figure}

The main conclusion of this result is that, as proved analytically,
the time and space GDE accounts for the three different diffusive
regimes that can be observed in the diffusion of a protein on DNA:
subdiffusion, normal diffusion and superdiffusion. In Fig. \ref{example}
we give an example of the time evolution of the diffusion of a particle
along a linear chain of $N=1001$ nodes. We use $\alpha=0.5$, $s=2.5$,
which according to the contour plot in Fig. \ref{fig:contour}, corresponds
to the subdiffusive regime. The value of $\gamma$ calculated from
the slope of the curve in Fig. \ref{example} (b) is $\gamma\approx0.33\pm0.01$
and according to the one of Fig. \ref{example} (c) is $\gamma\approx0.31\pm0.02$.
The analytical value is $\gamma=1/3$.

\begin{figure}
\subfloat[]{\begin{centering}
\includegraphics[width=0.5\textwidth]{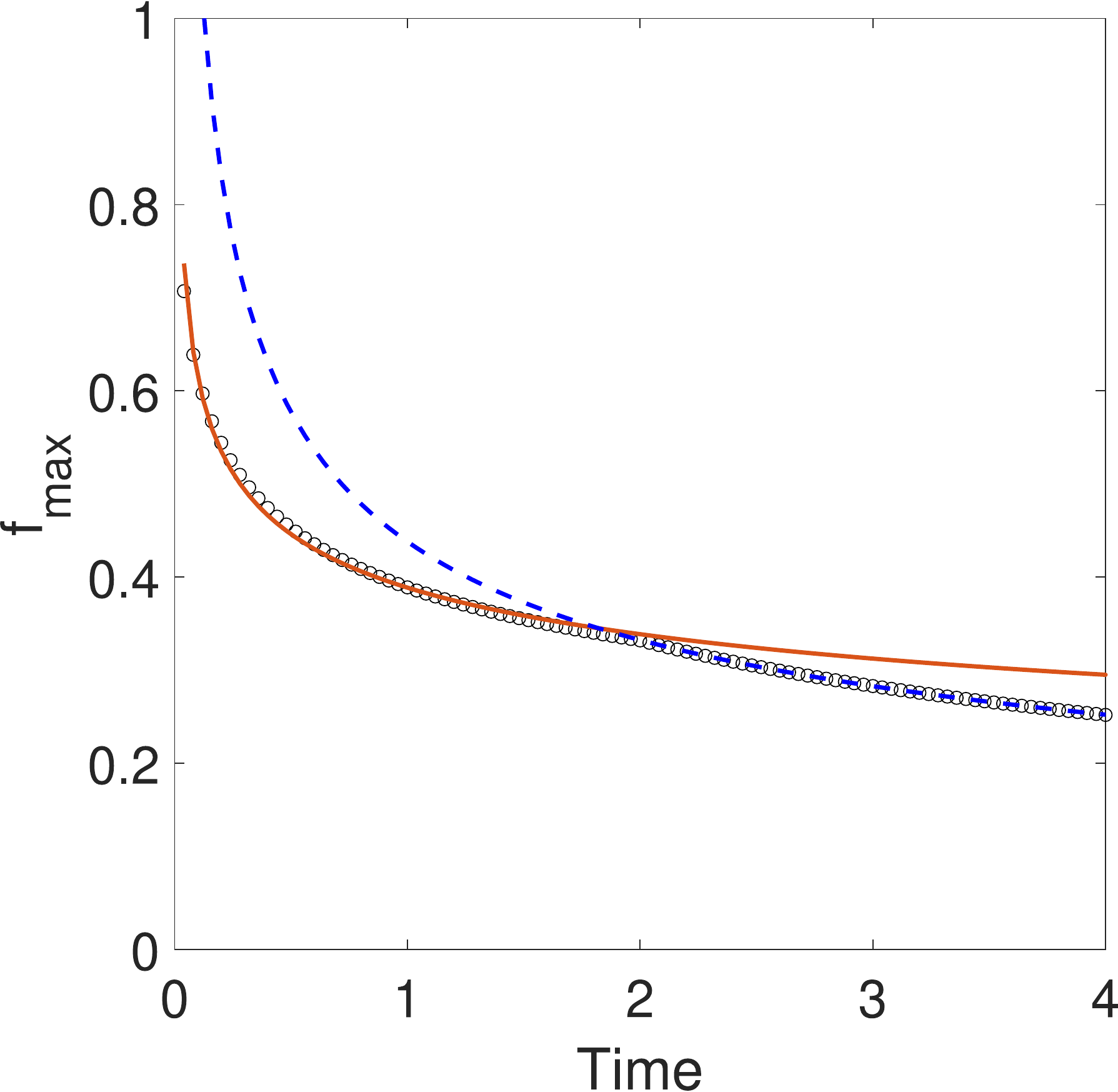}
\par\end{centering}

}\subfloat[]{\begin{centering}
\includegraphics[width=0.5\textwidth]{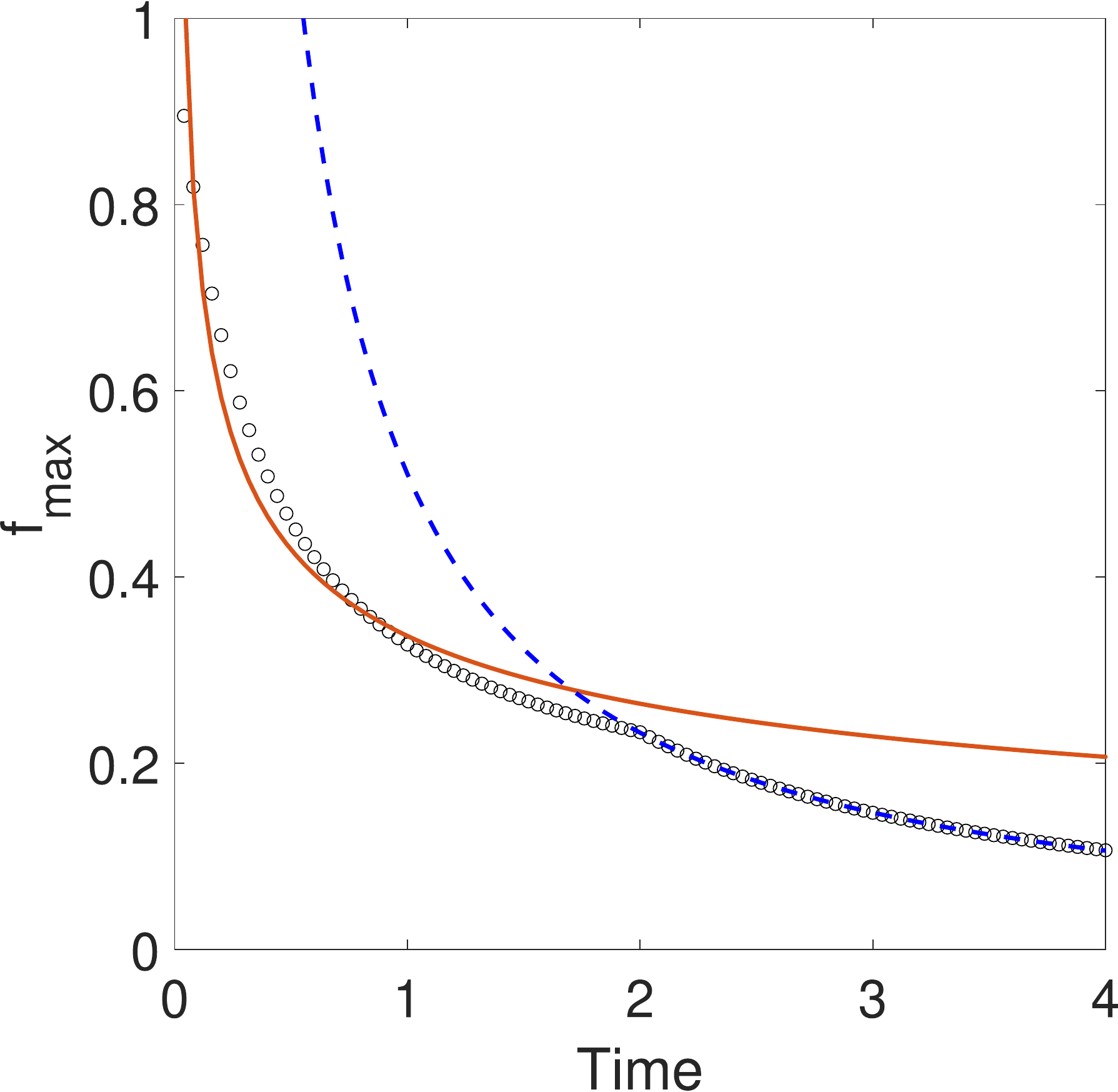}
\par\end{centering}
}

\caption{Decay of $f_{\textnormal{max}}$ for the diffusion of a particle along
a path graph of $N=1001$ nodes using the GDE with periodic temporal alternancy, with $s_{1}=200$,
$s_{2}=2$ and $\alpha=0.5$ (panel (a)) and $\alpha=0.9$ (panel
(b)). The solid (red) curve is the best power-law fitting for the
$f_{\textnormal{max}}$ vs. $t$ for $0<t<2$, and the broken (blue)
line is the same for $2<t<4$ (see text for details).}

\label{temporal_1}
\end{figure}

We now turn our attention to the study of the time-varying GDE with
periodic variation of the parameter $s$. We consider that $s$
oscillates between the values $s_{1}=200$, and $s_{2}=2$. That is,
for a given values of $\alpha$ the particle starts diffusing without
long-jumps ($s_{1}=200$). Then, after half a period of time it swaps to
a regime where long-jumps are allowed ($s_{2}=2$), and the process
is repeated cyclically. We first fix $\alpha=0.5$ and obtain the
results illustrated in Fig. \ref{temporal_1} (a). For $0<t<2$ we
have $f_{\textnormal{max}}(t)\approx0.3887t^{-0.1987}$ and for $2<t<4$
we obtain $f_{\textnormal{max}}(t)\approx0.4381t^{-0.3982}.$ That
is, when $\alpha=0.5$ the diffusive particle oscillates between two
subdiffusive regimes. In fact, the global fit of the process is given
by: $f_{\textnormal{max}}(t)\approx0.3671t^{-0.2472}$, which is a
clear signature of subdiffusion.

We now consider the case where $s_{1}=200$ and $s_{2}=2$ as before,
but using $\alpha=0.9$ as illustrated in Fig. \ref{temporal_1} (b).
Here, $f_{\textnormal{max}}(t)\approx0.3368t^{-0.3513}$ for $0<t<2$
and $f_{\textnormal{max}}(t)\approx0.5108t^{-1.133}$ for $2<t<4$.
The global process has scaling $f_{\textnormal{max}}(t)\approx0.2715t^{-0.4651}$.
That is, the global process is a subdiffusive regime, although the
diffusive particle alternates between a subdiffusive regime, i.e.,
$\gamma\approx-0.3513$ for $0<t<2$ , and a superdiffusive motion,
$\gamma\approx-1.133$ for $2<t<4$. 

\begin{figure}
\begin{centering}
\includegraphics[width=0.6\textwidth]{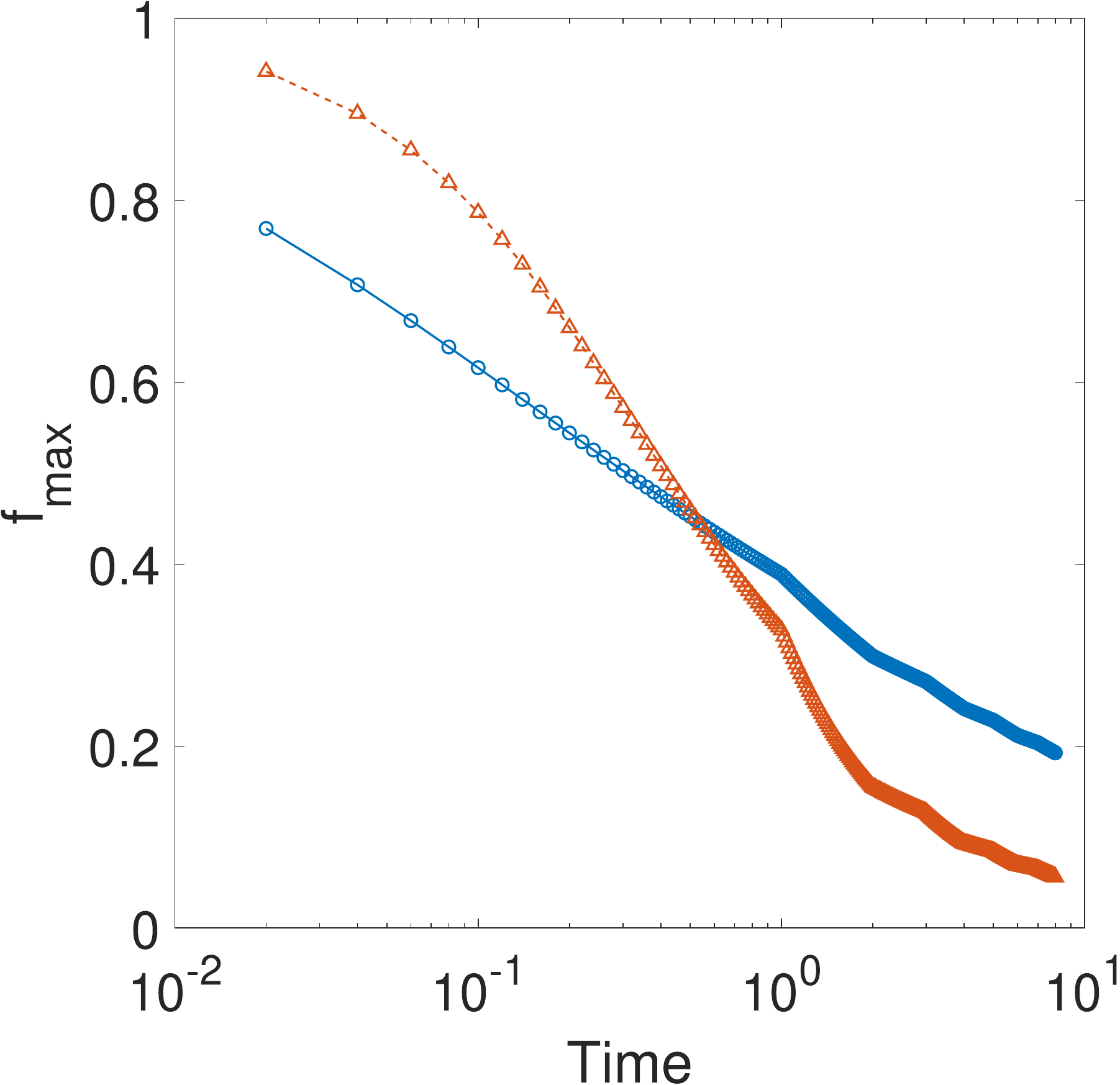}
\par\end{centering}
\caption{Decay of $f_{\textnormal{max}}$ for the diffusion of a particle along
a path graph of $N=1001$ nodes using the GDE with periodic temporal alternancy, with $s_{1}=200$,
$s_{2}=2$ and $\alpha=0.5$ (blue circles) and $\alpha=0.9$ (red
triangles). The solid lines are used to guide the eye. The $x$-axis
is in logarithmic scale to maximize the visualization effects.}

\label{temporal_2}
\end{figure}

The importance of this difference between the local temporal scale
and the global one is revealed when we plot the decay of $f_{\textnormal{max}}$
vs. $t$ for both values of the fractional parameter $\alpha$ as
illustrated in Fig. \ref{temporal_2}. As can be seen, the process
where $\alpha=0.5$ (blue circles) goes initially much faster than
that where $\alpha=0.9$ (red triangles). Then, there is a time in
which the process with $\alpha=0.9$ is much faster than that with
$\alpha=0.5$. In other words, the combination of subdiffusion with
superdiffusion, like when $\alpha=0.9$, allows the particle to make
a slower initial exploration of a region of the linear chain in comparison
with a subdiffusive-subdiffusive exploration. Additionally, the subdiffusive-superdiffusive
process produces a faster global convergence of the process due to
the long-jumps occurring in the superdiffusive regime. Notice that
the subdifusive-subdifusive regime is obtained when the fractional
parameter is relatively small, which correspond to systems with relatively
large temporal memory. However, the subdiffusive-superdiffusive alternancy
is obtained when the temporal memory is relatively small. Translating
these results to the case of a protein diffusing along a DNA chain
they mean that the alternant combination of sliding with jumping and/or
intersegmental transfer offer some important advances to the exploration
of the DNA by the protein. In this case, the slow subdiffusive regime
allows a detailed exploration of small DNA regions to find potential
targets and the fast superdiffusive regimes allow an exploration of
vast regions of the DNA chain in relatively short times. 

\section{Conclusions and Outlook}

We have defined a time and space GDE on undirected graphs/networks.
It uses a combination of fractional derivatives and Mellin-transformed $d$-path
Laplacian operator. We have found analytically the solution of this
equation and obtained the regions of the parametric space for which
an infinite one-dimensional system displays, normal, sub- and superdiffusion.
We have illustrated how this GDE can be applied to the study of the
diffusion of proteins along the one-dimensional structure of DNA,
where the mechanisms of sliding, hopping and intersegmental transfer,
may give rise to normal, sub- and superdiffusive behaviors. We have
also considered a GDE in which the parameters of the model change
with time allowing the temporal alternancy of the normal and anomalous
diffusive regimes. 

The current model is useful for any discrete system in which any combination
of the normal and anomalous diffusive regimes exists. The extension
of this model to consider directed graphs, multigraphs and simplicial
complexes can be performed to extend the areas of application of this
approach. Also important should be the analysis of networks beyond
the one-dimensional case presented here and to consider the influence
of network topologies, e.g., small-worldness, scale-freeness, etc.,
on the diffusive dynamics. All in all we consider that this GDE will
open new research avenues in the study of dynamical processes on graphs/networks.

\section*{Acknowledgement}

FD-D and EE acknowledge support from the Spanish Agency of Research (AEI) through Maria de Maeztu Program for units of Excellence in R\&D (MDM-2017-0711). FD-D thanks finantial support MDM-2017-0711-20-2 funded by MCIN/AEI/10.13039/50110 0 011033 and by FSE invierte en tu futuro. EE thanks Grant PID2019-107603GB-I00 by MCIN/AEI /10.13039/50110 0 011033.

\bibliographystyle{ieeetr}
\bibliography{bibliography_anomalous_diff}

\end{document}